\documentclass{aa}
\usepackage{graphics,epsfig,txfonts,color}
\usepackage[dvips]{rotating}
\usepackage{natbib}
\bibpunct{(}{)}{;}{a}{}{,}

\newcommand{\ergcm}[1]{$\times 10^{#1}$ erg cm$^{-2}$ s$^{-1}$}

\newcommand{\ergs}[1]{$\times 10^{#1}$ erg s$^{-1}$}

\newcommand{\hcm}[1]{$\times 10^{#1}$ cm$^{-2}$}

\newcommand{\nh}{$N_\mathrm{H}$}
\newcommand{\rh}{$r_\mathrm{h}$}

\newcommand{\ltsima}{$\buildrel < \over \sim$}
\newcommand{\lsim}{\lower.5ex\hbox{\ltsima}}
\newcommand{\gtsima}{$\buildrel > \over \sim$}
\newcommand{\gsim}{\lower.5ex\hbox{\gtsima}}

\newcommand{\hess}{H.E.S.S.}

\newcommand{\chandra}{\emph{Chandra}}

\newcommand{\fermi}{\emph{Fermi}}

\newcommand{\terzan}{Terzan~5}
\begin{document}
\definecolor{orange}{cmyk}{0,.5,1,0}
 
\title{A search for diffuse X-ray emission from GeV-detected Galactic globular clusters}
\author{P.\,Eger\inst{1}, and
		W.\,Domainko\inst{2}
	   }

\titlerunning{A search for diffuse X-ray emission from GeV detected GCs}
\authorrunning{Eger et al.}

\institute{\inst{1} Erlangen Centre for Astroparticle Physics, 
				Universit\"at Erlangen-N\"urnberg, Erwin-Rommel-Str. 1, 91058 Erlangen, Germany\\
		   \inst{2} Max-Planck-Institut f\"ur Kernphysik, PO Box 103980, 69029 Heidelberg, Germany\\
           \email{peter.eger@physik.uni-erlangen.de}
           }
 
\date{Received / Accepted }
 
\abstract{
Recently, diffuse and extended sources in TeV $\gamma$-rays as well as in X-rays have been detected in the direction of the Galactic globular cluster (GC) \terzan . Remarkably, this is among the 
brightest GCs detected in the GeV regime. 
The nature of neither the TeV nor the diffuse X-ray signal from \terzan\ is not yet settled. 
These emissions most likely indicate the presence of several non-thermal radiation processes in
addition to those that give rise to the GeV signal.
} 
{
The aim of this work is to search for diffuse X-ray emission from all GeV-detected GCs where appropriate X-ray observations are available, and to compare the obtained results with the signal detected from \terzan . 
This study will help to determine whether Terzan 5 stands out among other GC or whether a whole population of globular clusters feature similar properties.
}
{
After assessing all archival X-ray observations of GeV detected GCs, we analyzed the data of six \chandra\ observations pointed toward M\,62, NGC\,6388, NGC\,6541, M\,28, M\,80, and NGC\,6139. 
For each GC we characterized the diffuse X-ray emission using the same analysis techniques as for \terzan . 
To study the emission on the same relative scales we used the half-mass radius as a scale parameter to determine the extent of the potential emission region.
}
{
None of the six GCs show significant diffuse X-ray emission above
the particle and diffuse galactic X-ray background components. 
The derived upper limits allow to assess the validity of different models that were discussed in the interpretation of the multi-wavelength data of \terzan . 
A scenario based on synchrotron emission from relativistic leptons provided by the millisecond pulsar population cannot be securely rejected if a comparable magnetic field strength as in \terzan\ is assumed for every GC. 
However, such a scenario seems to be unlikely for NGC\,6388, and M\,62. 
An inverse-Compton scenario relying on the presence of a putative GRB remnant with the same properties as proposed for \terzan\ can be ruled out for all six GCs. 
Finally, the assumption that each GC hosts a source with the same luminosity as in \terzan\ is ruled out for all GCs but NGC\,6139.
}
{}

 
\maketitle

\section{Introduction}
\label{sec-introduction}
Globular clusters (GCs) are dense systems of very old stellar populations. 
Owing to the extreme densities in their cores, GCs feature very high stellar encounter rates that lead to high probabilities for the dynamical formation of compact binaries \citep{2006ApJ...646L.143P,2003ApJ...591L.131P,2010ApJ...714.1149H}. 
This in turn likely gives rise to the many of millisecond pulsars (msPSRs) that have been discovered in some of these systems \citep{2005ASPC..328..147C,2008IAUS..246..291R}. 

Recently, several Galactic GCs could be associated to GeV $\gamma$-ray sources detected with the \fermi /LAT observatory \citep{2009Sci...325..845A,2010A&A...524A..75A,2010ApJ...712L..36K,2011ApJ...729...90T}. 
The $\gamma$-ray spectra of most of these sources feature cut-offs at a few GeV, very reminiscent of the spectra measured from individual msPSRs \citep[see, e.g.,][]{2009ApJ...699L.102A,2009Sci...325..848A}. 
Indeed, an individual msPSR in a GC has been detected recently in the GeV regime with \fermi\ \citep[see][]{2011arXiv1111.3754T}. 
This is one of the reasons why the GeV $\gamma$-ray signals from GCs are generally interpreted as the cumulative magnetospheric emission from their whole msPSR population. 
However, the observed emission in the GeV regime could also originate from inverse Compton (IC) up-scattering of ambient photon fields by leptons accelerated in pulsar wind nebulae (PWNe) surrounding individual msPSRs \citep{2010ApJ...723.1219C,2011ApJ...726..100H}. 
In certain cases the resulting spectra would look very similar in the GeV range to those observed by \fermi\ from these GCs.

\begin{table*}[t]
\caption[]{The GC sample and \chandra\ observations}
\begin{center}
\begin{tabular}{lllllll}
\hline\hline\noalign{\smallskip}
\multicolumn{1}{l}{GC} &
\multicolumn{1}{l}{Reference$^{(1)}$} &
\multicolumn{1}{l}{$r_\mathrm{h}^{(2)}$} &
\multicolumn{1}{l}{ObsID$^{(3)}$} &
\multicolumn{1}{l}{Coverage$^{(4)}$} & 
\multicolumn{1}{l}{Net exposure$^{(5)}$} & 
\multicolumn{1}{l}{Sensitivity$^{(6)}$} \\ 
\multicolumn{1}{l}{name} &
\multicolumn{1}{l}{} &
\multicolumn{1}{l}{($^{\prime\prime}$)} &
\multicolumn{1}{l}{} &
\multicolumn{1}{l}{(\%)} &
\multicolumn{1}{l}{(ks)} &
\multicolumn{1}{l}{($10^{-15}$ erg cm$^{-2}$ s$^{-1}$)} \\
\noalign{\smallskip}\hline\noalign{\smallskip}
M 62			& A10		& 74 	& 2677 (ACIS-S)				& 98		& 62.2	& 0.6\\ 
NGC 6388		& A10		& 40	& 5505 (ACIS-S)				& 99	 	& 44.7	& 1.0\\ 
NGC 6541		& A10,T11	& 71	& 3779 (ACIS-S)				& 80		& 44.9	& 0.7\\ 
M 28			& A10		& 94	& 9132 (ACIS-S)				& 63		& 141.9 & 0.5\\ 
M 80			& T11		& 37	& 1007 (ACIS-S)				& 99	 	& 48.6	& 1.0\\ 
NGC 6139		& T11		& 51	& 8965 (ACIS-S)				& 83		& 18.1	& 2.0\\ 
\noalign{\smallskip}\hline\noalign{\smallskip}
47 Tuc			& A09,A10	& 190	& $r_\mathrm{h}$ too large	& 			&		& \\
$\omega$ Cen	& A10		& 134	& $r_\mathrm{h}$ too large	& 			&		& \\
NGC 6440		& A10		& 35	& bright point source		& 			&		& \\
NGC 6441		& A10,T11	& 34	& bright point source		& 			&		& \\
NGC 6624		& T11		& 49	& only grating or low exposure			& 		&	&	\\
NGC 6652		& A10		& 29	& only low exposure 		& 			&	&	\\
M 15			& A10		& 60	& bright point source		& 			&	&	\\
NGC 6752		& A10,T11	& 115	& \multicolumn{3}{l}{$r_\mathrm{h}$ large, known diffuse source in region \citep{2007PASJ...59..727O}} \\
Liller 1		& T11		& 48	& only grating				& 			&	&	\\
\hline\noalign{\smallskip}
\end{tabular}
\label{tab-observations}
\end{center}
Summary of GeV-bright GCs that were considered (top section) or not considered (bottom section) in this work. 
$^{(1)}$\fermi /LAT reference: 
A09 \citep{2009Sci...325..845A}, 
A10 \citep{2010A&A...524A..75A},
T11 \citep{2011ApJ...729...90T}
$^{(2)}$half-mass-radius (\rh ) from \citet{1996AJ....112.1487H}. 
$^{(3)}$\chandra\ observation identifier and instrument mode, or reason for exclusion. 
$^{(4)}$Fraction of the area between one and three \rh\ within the FoV. 
$^{(5)}$Remaining exposure after GTI screening. Owing to low background activity this is identical to the total exposure in most cases (see Sect.~\ref{sec-preparation-point-sources}). 
$^{(6)}$Sensitivity for point-source detection (0.5--7.0~keV). 
\end{table*}

Furthermore, a TeV $\gamma$-ray source was recently detected with \hess\ in the vicinity of \terzan\ \citep{2011A&A...531L..18H}. 
Intriguingly, this GC is among the brightest GCs seen with \fermi\ \citep{2010A&A...524A..75A} and also the one with the most detected msPSRs (34\footnote{http://www.naic.edu/$\sim$pfreire/GCpsr.html (and references therein)}).
In the current literature, two models predict TeV $\gamma$-ray emission roughly at a level detected from \terzan . 
Similar to what was also suggested for the GeV range (see previous paragraph), these models both involve  IC up-scattering of ambient photon fields, such as starlight from GC member stars and the cosmic microwave background, by relativistic leptons. 
As in the GeV regime, these leptons could be accelerated either in individual PWNe surrounding msPSRs \citep{2007MNRAS.377..920B,2009ApJ...696L..52V} or even be re-accelerated in colliding PWNe shocks \citep{2010ApJ...723.1219C}. 

Leptonic models for TeV $\gamma$-ray emission from GCs also predict X-ray emission caused by synchrotron radiation from the same lepton population \citep{2008AIPC.1085..277V}. 
Indeed, diffuse X-ray emission centered on the core and extending well beyond the half-mass-radius (\rh ) has recently been detected with \chandra\ from \terzan\ \citep{2010A&A...513A..66E}. 
The spectrum of this emission is well represented by an absorbed powerlaw model with a spectral index of 0.9$\pm$0.5 and therefore is most likely of non-thermal origin. 
Diffuse X-ray emission has been measured from a number of other Galactic GCs \citep[see, e.g.,][]{2007PASJ...59..727O}. 
However, in these cases the spectra were softer and the emission region was either offset from the core in the direction of the  GCs proper motion or featured an arc-like morphology. 
Here, the emission was interpreted as either non-thermal bremsstrahlung from shock-accelerated electrons hitting nearby gas clouds, or thermal emission from shock-heated gas \citep{2007PASJ...59..727O}. 
Following \citet{2010A&A...513A..66E}, these scenarios are most likely not valid for \terzan , whereas synchrotron radiation appears to be a plausible explanation considering the energetics.
Based on a study of available multi-wavelength data of \terzan , predominantly in the radio regime, \citet{2011A&A...532A..47C} were able to disfavor non-thermal bremsstrahlung, while still leaving 
other non-thermal processes, such as IC or synchrotron radiation, as valid scenarios. 
Following \citet{2010ApJ...723.1219C}, the diffuse X-ray signal could be interpreted as the tail of the IC peak at TeV energies. 
However, these authors' prediction for the flux is a factor of 3--4 lower than what is observed from \terzan . 

Another possible explanation for the TeV gamma-ray emission is related to a remnant of a short gamma-ray burst (GRB) \citep{2011A&A...533L...5D} resulting from a neutron star--neutron star merger \citep{2006NatPh...2..116G}. 
Owing to their high percentage of stellar binaries, GCs are prime candidates to host these events because they form considerable numbers of potential progenitors.
Here the $\gamma$-rays could be produced in inelastic collisions of hadrons, accelerated in the shocks, with ambient interstellar matter. 
Then the TeV $\gamma$-ray signal could be accompanied by thermal X-ray emission arising from shock-heated gas \citep{Domainko:2008fja} or non-thermal IC radiation from primary leptons accelerated along with the hadrons in the shocks \citep{2011A&A...533L...5D}. 
Particularly for the latter case the TeV and X-ray signals that have been detected from \terzan\ could be related. 

The aim of this work is to systematically search for diffuse hard X-ray emission from other Galactic GCs that have been detected in the GeV energy range by \fermi .
Based on different model assumptions we then investigate whether this kind of emission could be typical for a whole population of GCs or if \terzan\ may stand out, as this GC would do if there were a recent and rare catastrophic event. 
To facilitate a comparison of the obtained results to \terzan , we studied the diffuse emission on similar relative scales, using \rh\ as a reference parameter to determine the size of each potential emission region.

\section{X-ray observations, analysis and results}
\label{sec-analysis}
\subsection{The globular cluster sample and \chandra\ observations}
\label{sec-gc-sample}
We built the list of GeV-bright GCs detected by \fermi\ based on the systematic studies performed by \citet{2010A&A...524A..75A} and \citet{2011ApJ...729...90T}, regardless of the measured spectral properties and their interpretation. 
Furthermore, we restricted our study to \chandra\ data \citep{2002PASP..114....1W} because it is the only X-ray telescope that can cope with the task of effectively removing point-like X-ray sources in crowded areas close to GCs. 
The narrow point spread function (PSF) of \chandra '\emph{s} telescope keeps the potential contamination of regions outside the removed point sources to a minimum. 
Fortunately, all selected GCs have been observed by \chandra\ and the data were accessible via the public archive. 
To retain as much spatial and spectral information as possible at the same time, we only considered the non-grating \emph{Advanced CCD Imaging Spectrometer} \citep[ACIS,][]{2003SPIE.4851...28G} data with an exposure of at least 10~ks. 

To judge which of these objects are suited for our study we required that more than 50\% of the region of a potential diffuse X-ray signal is covered by the ACIS field of view (FoV). 
For \terzan\ the significant diffuse excess emission lies between one and three \rh , a scale we used to define the coverage fraction for all other GCs. 
The respective values for \rh\ were taken from \citet{1996AJ....112.1487H}. 
Furthermore, we required that no X-ray point source brighter than 1\ergcm{-12} (see Sect.~\ref{sec-preparation-point-sources}) is located within the FoV to avoid contamination from the PSF wings in our region of interest. 
For NGC\,6752 and M\,80, already known diffuse X-ray sources are located within 3$\times r_\mathrm{h}$ \citep{2007PASJ...59..727O}, which would have to be removed from the data to search for an additional signal. 
However, for NGC\,6752 the area of the remaining region is below the required minimum coverage fraction of 50\% and this GC was therefore excluded from the analysis. 
The sample of GeV-bright GCs is compiled in Table~\ref{tab-observations} where the top section lists all GCs passing our \emph{a priori} cuts and the bottom section gives short comments on why the respective GC was excluded from our study. 

\subsection{Data preparation and point-source detection}
\label{sec-preparation-point-sources}
All steps of the analysis procedure were performed in the same way as for our study of \terzan\ \citep[see][]{2010A&A...513A..66E}. 
For the X-ray analysis we used the CIAO software version 4.3, supported by tools from the FTOOLS package \citep{1995ASPC...77..367B} and XSPEC version 12.5.0 for spectral modeling \citep{1996ASPC..101...17A}. 
The \texttt{event1} data were reprocessed with the latest position- and energy calibration (CTI correction, v4.4.6) using bad pixel files generated by \texttt{acis\_run\_hotpix}. 
To screen the ACIS-S data from periods of increased background flaring activity, we used the good-time-interval (GTI) file that is provided by the standard processing chain. 
In most of the observations the background level was fairly low, and the net exposures after the screening are virtually identical to the total lifetime. 
To test the standard GTI filtering we inspected the light curves created from events in the 9--12\,keV band for each observations. 
However, no additions to the standard GTIs were necessary for any of the six datasets. 
The resulting net exposures are also listed in Table~\ref{tab-observations}.

To detect and remove point-like X-ray sources, we ran the wavelet detection algorithm \texttt{wavdetect} in the three energy bands 0.5--2~keV, 2--7~keV, and 0.5--7~keV. 
These bands were chosen to be sensitive to soft Galactic sources, such as active stars, hard sources associated to the GCs, and absorbed background sources, e.g. AGN. 
The sensitivities for the detection of point sources are also given in Table~\ref{tab-observations}.
To estimate whether the detected sources are predominantly related to the GC or belong to a population of background sources, such as AGN, we used the results from \citet{2001ApJ...551..624G} to calculate the estimated number density of background sources based on the detection sensitivity of the individual observations. 
We found that the number densities of expected background sources lie between $\sim$1900 sources~deg$^{-2}$ (M~62) and 675 sources~deg$^{-2}$ (NGC~6139). 
For an individual ACIS CCD chip this yields total numbers of expected sources of 34 (M~62) and 12 (NGC~6139), respectively. 
These results are compatible with similar calculations for \terzan\ made by \citet{2006ApJ...651.1098H}, who expected a total number of eight AGN per CCD chip above a sensitivity threshold of $\sim$3\ergcm{-15} ($\approx$10 counts). 
We note that the galactic column densities in the directions of the GCs considered here are about an order of magnitude lower than toward \terzan . 
Therefore, we did not correct the number of expected background sources for absorption effects. 
Except for the region within \rh\ these estimated number densities of background sources agree very well with the number of detected sources, implying that most of these sources are of extra-galactic origin. 

To estimate the fluxes of these sources based on their count-rate, we assumed a typical non-thermal spectrum, such as measured for X-ray sources in \terzan\ \citep{2006ApJ...651.1098H}. 
Here we assumed an intrinsic powerlaw spectrum with index $-$2 and took into account the total Galactic column density \citep[\nh , from][]{1990ARA&A..28..215D} in the respective direction. 
Among all considered GCs the values of \nh\ are moderate and on the order of a few times $10^{21}$~cm$^{-2}$. 
To estimate the flux, we first corrected the count-rates for the effects of the mirror vignetting based on the off-axis angle at each source position. 
If the derived flux of any of the point sources in the FoV was above a level of 1\ergcm{-12}, the respective GC was excluded from the sample (see Sect.~\ref{sec-gc-sample}). 
For the remaining six GCs we removed all detected point sources from the data using the 3\,$\sigma$ width of the PSF at the source position. 

As already mentioned in Sect.~\ref{sec-gc-sample}, an already known diffuse X-ray source is located close to M\,80 \citep[see][]{2007PASJ...59..727O}. 
This source is located toward the southwest of the GC center and the significant emission extends out to $\sim$4\rh . 
To remove this source from the data we excluded a circle segment with an opening angle of 90$^\circ$ oriented toward the southwest (270$^\circ$--360$^\circ$, west to north) and extending out to 5\rh . 

\subsection{Diffuse emission}
\label{sec-diffuse emission}
To characterize the level and radial dependence of potential diffuse X-ray emission from each of the six GCs on the same relative scales as for \terzan , we extracted spectra from the same eight annular regions (``Ring1" to ``Ring8") with inner and outer radii in units of \rh . 
Because the X-ray point source populations of the GCs are much smaller compared to \terzan , we also considered regions closer than 1$\times$\rh\ that had no high contamination from unresolved point sources. 
Therefore, we introduced one additional annulus (``Ring0") closer toward the GC center with the same widths as the other eight annuli. 
We chose a scaling of the region sizes with respect to \rh\ because it reflects the typical size of the potential emission region for all scenarios. 
For IC emission from relativistic leptons the emission is expected to follow the profile of the target photon field, whereas in an msPSRs scenario the emission should reflect the distribution of sources within the GC. 
We used the values for the position and \rh\ as given in the GC catalog of \citet{1996AJ....112.1487H}. 
Table~\ref{tab-regions} lists the details of these regions. 
The mean effective area (ARF) and energy resolution (RMF) files were calculated by averaging over all contributing pixels assuming a flat detector map. 
The ARFs and RMFs created using a weighted average over all pixels with weights corresponding to acceptance-corrected count-rates yielded comparable results. 

While the position of the GC core is always located in the area covered by the back-illuminated ACIS-S3 chip, in some cases parts or all of outer three annuli were on one of the two neighboring front-illuminated chips. 
In these cases we extracted the spectrum and response files for each CCD separately to fit the datasets in parallel (see next paragraph). 
As for \terzan\ \citep[see][]{2010A&A...513A..66E}, to estimate the non-X-ray background (NXB) component we used a background dataset from the calibration database where the ACIS detector was operated in its stowed position. 
For each spectrum from the annular regions a corresponding background spectrum was extracted from the NXB dataset also using the identical point-source exclusion regions. 
To account for the time-dependence of the NXB, we scaled the background by the ratio 
of the source and background count-rates in the 9--12~keV energy band for each spectrum 
\citep[as described by][]{2003ApJ...583...70M}. 

\begin{table}
\caption[]{Parameters of the annular extraction regions}
\begin{center}
\begin{tabular}{l|llllll}
\hline\hline\noalign{\smallskip}
\multicolumn{1}{l}{} &
\multicolumn{1}{l}{M 62} &
\multicolumn{1}{l}{NGC} &
\multicolumn{1}{l}{NGC} &
\multicolumn{1}{l}{M 28} &
\multicolumn{1}{l}{M 80} &
\multicolumn{1}{l}{NGC} \\
\multicolumn{1}{l}{} &
\multicolumn{1}{l}{} &
\multicolumn{1}{l}{6388} &
\multicolumn{1}{l}{6541} &
\multicolumn{1}{l}{} &
\multicolumn{1}{l}{} &
\multicolumn{1}{l}{6139} \\
\noalign{\smallskip}\hline\noalign{\smallskip}
width$^{(1)}$	& 36	& 19	& 34	& 45	& 18	& 25	\\ 
radius0$^{(2)}$ & 46	& 25	& 44	& 58	& 23	& 32	\\
radius1			& 81	& 44	& 78	& 103	& 40	& 56	\\ 
radius2			& 117	& 63	& 112	& 148	& 58	& 81	\\ 
radius3			& 152	& 82	& 146	& 193	& 75	& 105	\\ 
radius4			& 186	& 101	& 179	& 236	& 92	& 129	\\ 
radius5			& 222	& 120	& 213	& 281	& 110	& 153	\\ 
radius6			& 258	& 139	& 247	& 326	& 127	& 177	\\ 
radius7			& 293	& 158	& 281	& 371	& 145	& 202	\\ 
radius8			& 329	& 178	& 315	& 416	& 163	& 226	\\ 
\hline\noalign{\smallskip}
\end{tabular}
\label{tab-regions}
\end{center}
All values are given in arc seconds. 
$^{(1)}$Width of the annuli equals 0.48$\times$\rh\ for each GC. 
$^{(2)}$Inner radius of each ring calculated as (0.62, 1.10, 1.58, 2.06, 2.52, 3.00, 3.48, 4.00, 4.44) $\times$\rh . 
\end{table}

\begin{table}[ht]
\caption[]{Spectral fitting results}
\renewcommand{\tabcolsep}{4.pt}
\begin{center}
\begin{tabular}{lllll}
\hline\hline\noalign{\smallskip}
\multicolumn{1}{l}{Region} &
\multicolumn{1}{l}{Excess counts$^{(1)}$} &
\multicolumn{1}{l}{$\Gamma / kT ^{(2)}$} &
\multicolumn{1}{l}{Surface flux$^{(3)}$} &
\multicolumn{1}{l}{$\chi_\nu^2$ (d.o.f.)} \\
\noalign{\smallskip}\hline\noalign{\smallskip}
\multicolumn{5}{l}{\textbf{M 62}} \\
Ring0	  &	 $660.1 \pm	 43.3$	  & $8.9 \pm   0.3$	   & $2.51\pm 0.42$ & 1.2(504)	 \\
Ring1	  &  $897.4 \pm  49.5$    & $9.6 \pm   0.1$    & $2.18\pm 0.26$ & 		     \\
Ring2	  &  $1145.9\pm  48.1$    & $9.8 \pm   0.1$    & $2.05\pm 0.20$ &			  \\
Ring3	  &  $1175.2\pm  50.8$    & $9.9 \pm   0.1$    & $1.82\pm 0.12$ &			  \\
Ring4	  &  $1669.7\pm  58.3$    & $9.6 \pm   0.2$    & $2.22\pm 0.19$ &			  \\
Ring5	  &  $1750.0\pm  78.7$    & $9.7 \pm   0.1$    & $2.18\pm 0.18$ &			  \\
Ring6	  &  $850.1 \pm  41.8$    & $9.0 \pm   0.4$    & $1.91\pm 0.15$ &			  \\
Ring7	  &  $223.9 \pm  14.8$    & $8.6 \pm   0.5$    & $2.18\pm 0.25$ &			  \\
Ring8	  &  $326.5 \pm  17.7$    & $9.0 \pm   0.6$    & $2.13\pm 0.20$ &			  \\
Outer	  &	 $1672.5\pm  74.8$    & $0.40\pm  0.03$	   & $5.74\pm 0.37$ & 1.3(99)     \\
\hline\noalign{\smallskip}
\multicolumn{5}{l}{\textbf{NGC 6388}} \\
Ring0	  &  $264.9 \pm  26.5$    & $5.6 \pm   1.2$    & $6.65\pm 2.4 $ & 1.1(469)    \\
Ring1	  &  $318.0 \pm  37.8$    & $6.8 \pm   1.1$    & $3.79\pm 1.3 $ & 		      \\
Ring2	  &  $392.2 \pm  38.1$    & $7.0 \pm   0.9$    & $3.34\pm 0.91$ &			  \\
Ring3	  &  $446.8 \pm  37.0$    & $6.6 \pm   0.9$    & $3.72\pm 0.98$ &			  \\
Ring4	  &  $457.6 \pm  29.6$    & $7.1 \pm   0.9$    & $2.63\pm 0.66$ &			  \\
Ring5	  &  $580.4 \pm  46.2$    & $6.9 \pm   0.9$    & $3.03\pm 0.71$ &			  \\
Ring6	  &  $681.4 \pm  46.6$    & $7.4 \pm   0.9$    & $2.76\pm 0.61$ &			  \\
Ring7	  &  $705.5 \pm  55.4$    & $7.1 \pm   0.8$    & $3.20\pm 0.64$ &			  \\
Ring8	  &  $657.7 \pm  45.6$    & $7.3 \pm   0.9$    & $2.58\pm 0.59$ &			  \\
Outer	  &	 $1966.9\pm  76.5$ 	  & $0.18\pm  0.02$	   & $8.25\pm 0.55$ & 1.1(88)	  \\
\hline\noalign{\smallskip}
\multicolumn{5}{l}{\textbf{NGC 6541}} \\
Ring0	  &  $442.5 \pm  47.3$    & $8.9 \pm   0.3$    & $1.56\pm 0.19$ & 1.4(451)    \\
Ring1	  &  $730.2 \pm  53.8$    & $9.2 \pm   0.2$    & $1.42\pm 0.13$ & 		      \\
Ring2	  &  $917.3 \pm  63.4$    & $9.5 \pm   0.1$    & $1.34\pm 0.10$ &			  \\
Ring3	  &  $769.7 \pm  58.8$    & $9.2 \pm   0.2$    & $1.39\pm 0.12$ &			  \\
Ring4	  &  $437.4 \pm  21.0$    & $9.5 \pm   0.1$    & $1.20\pm 0.09$ &			  \\
Ring5	  &  $547.5 \pm  23.1$    & $9.1 \pm   0.1$    & $1.47\pm 0.10$ &			  \\
Ring6	  &  $450.1 \pm  21.6$    & $9.2 \pm   0.2$    & $1.38\pm 0.11$ &			  \\
Ring7	  &  $367.1 \pm  20.0$    & $9.4 \pm   0.2$    & $1.25\pm 0.11$ &			  \\
Ring8	  &  $345.0 \pm  18.9$    & $9.4 \pm   0.2$    & $1.25\pm 0.11$ &			  \\
Outer	  &  $1767.8 \pm  81.0$   & $0.17\pm  0.02$    & $5.51\pm 0.39$ & 0.9(67)     \\
\hline\noalign{\smallskip}
\multicolumn{5}{l}{\textbf{M 80}} \\
Ring0	  &  $187.6 \pm  24.6$    & $  2.5 \pm  1.6$   & $10.8\pm 8.0$ & 1.4(567)	  \\
Ring1	  &  $210.6 \pm  28.5$    & $  2.6 \pm  1.4$   & $6.9\pm 5.7$ & 			  \\
Ring2	  &  $291.6 \pm  28.4$    & $  3.3 \pm  0.6$   & $3.1\pm 1.5$ & 			  \\
Ring3	  &  $360.6 \pm  32.2$    & $  3.2 \pm  0.5$   & $3.9\pm 1.7$ & 			  \\
Ring4	  &  $432.2 \pm  35.1$    & $  1.6 \pm  0.9$   & $16 \pm 11 $ & 			  \\
Ring5	  &  $538.8 \pm  42.5$    & $  2.2 \pm  0.5$   & $8.1\pm 3.6$ & 			  \\
Ring6	  &  $626.3 \pm  43.2$    & $  2.0 \pm  0.6$   & $10 \pm 5.0$ & 			  \\
Ring7	  &  $673.4 \pm  51.3$    & $  2.5 \pm  0.4$   & $6.0\pm 2.2$ & 			  \\
Ring8	  &  $711.8 \pm  45.5$    & $  2.2 \pm  0.4$   & $9.4\pm 3.4$ & 			  \\
Outer	  &  $2309.4\pm  79.0$    & $  0.42\pm 0.08$   & $26.0\pm 2.5$ & 1.2(182)	      \\
\hline\noalign{\smallskip}
\end{tabular}
\label{tab-spectral-results}
\end{center}
The quoted uncertainties are 90\% confidence intervals calculated by letting each free parameter vary. 
Therefore, parameter correlations are taken into account.
$^{(1)}$Total number of excess counts summed over all contributing CCDs in the energy bands 1--7 keV (Ring1-8) and 0.7-10.0 keV (Outer). 
$^{(2)}$This column gives the photon index in the cases where an absorbed powerlaw spectrum was fitted (Ring1-8), or $kT$ in units of keV for the combined Outer region. 
$^{(3)}$Observed surface flux in units of 10$^{-8}$\,erg\,cm$^{-2}$\,s$^{-1}$\,sr$^{-1}$ in the energy bands 1--7 keV (Ring1-8) and 0.7-10.0 keV (Outer). 
\end{table}

\begin{figure*}
  \resizebox{0.98\hsize}{!}{\begin{turn}{-90}\includegraphics[clip=]{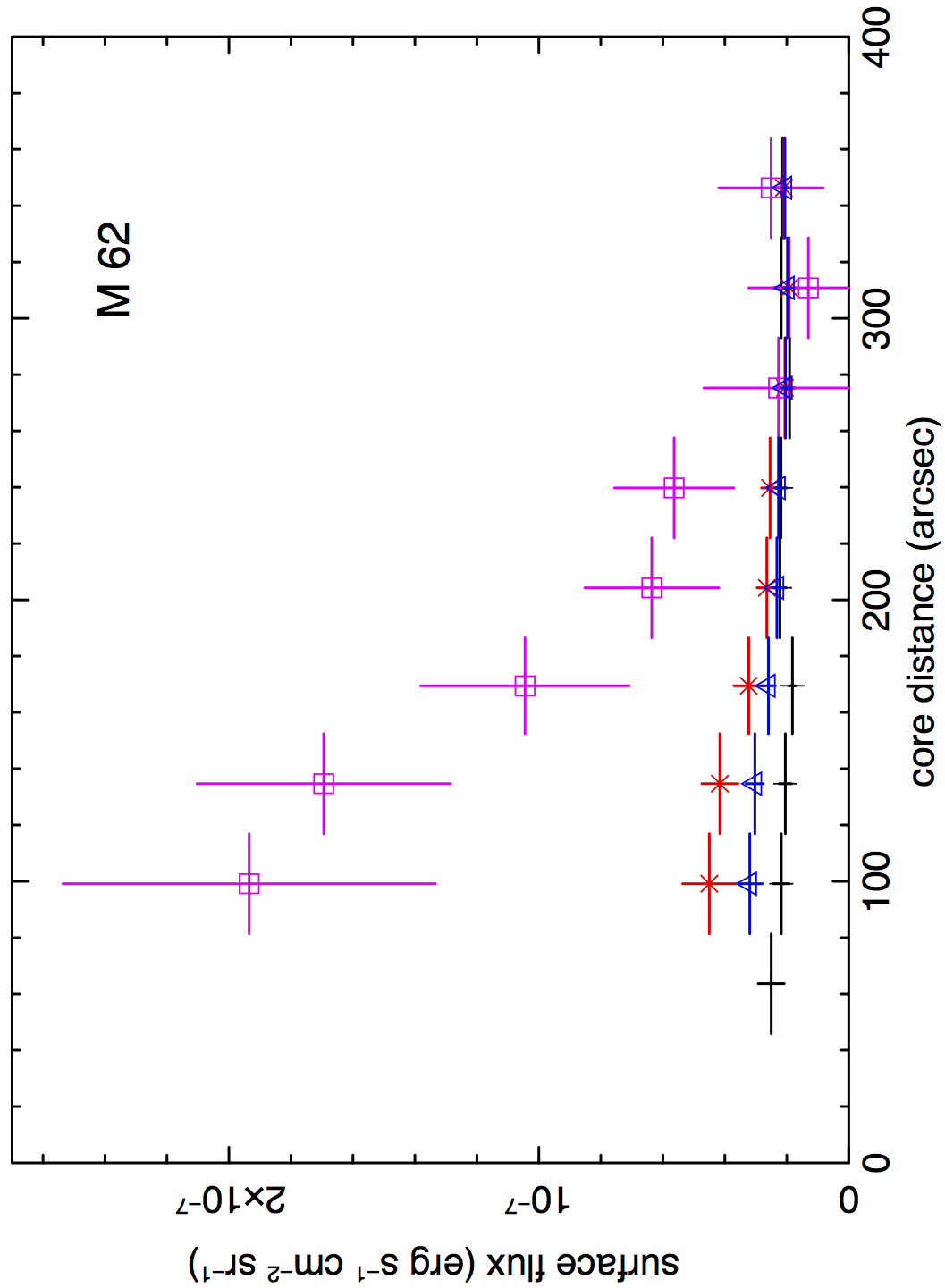}\end{turn}
  \begin{turn}{-90}\includegraphics[clip=]{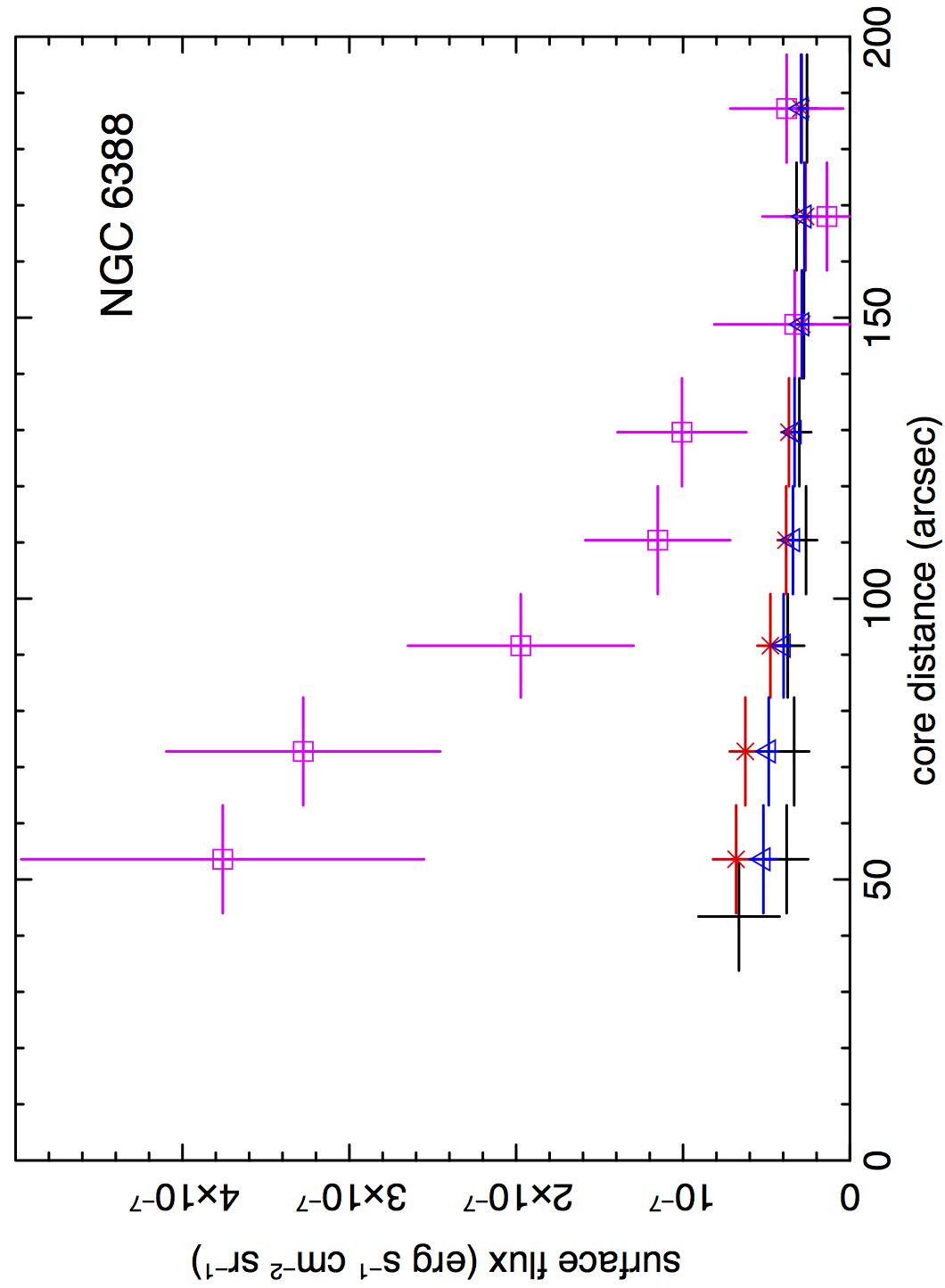}\end{turn}}  \resizebox{0.98\hsize}{!}{\begin{turn}{-90}\includegraphics[clip=]{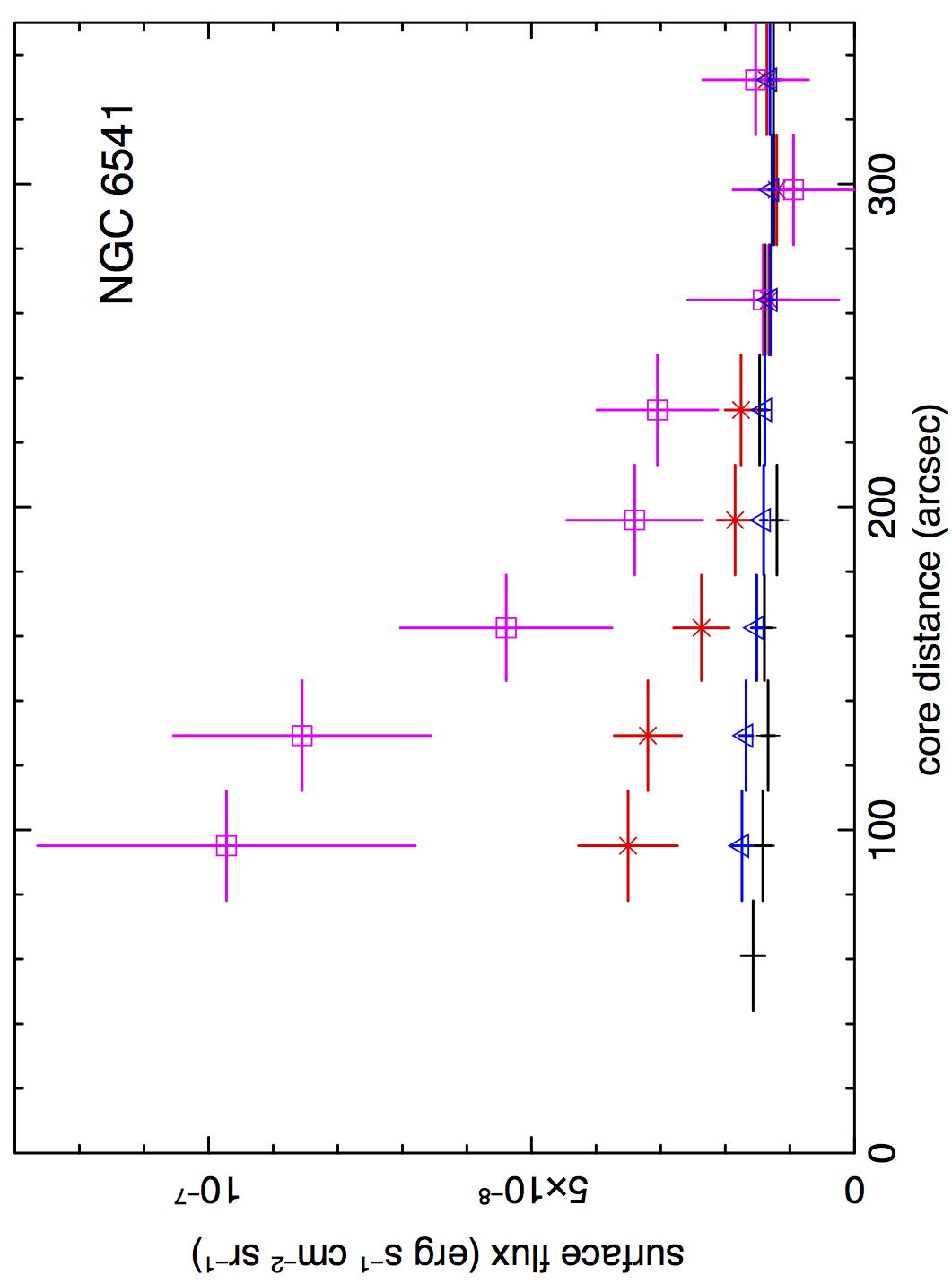}\end{turn}
  \begin{turn}{-90}\includegraphics[clip=]{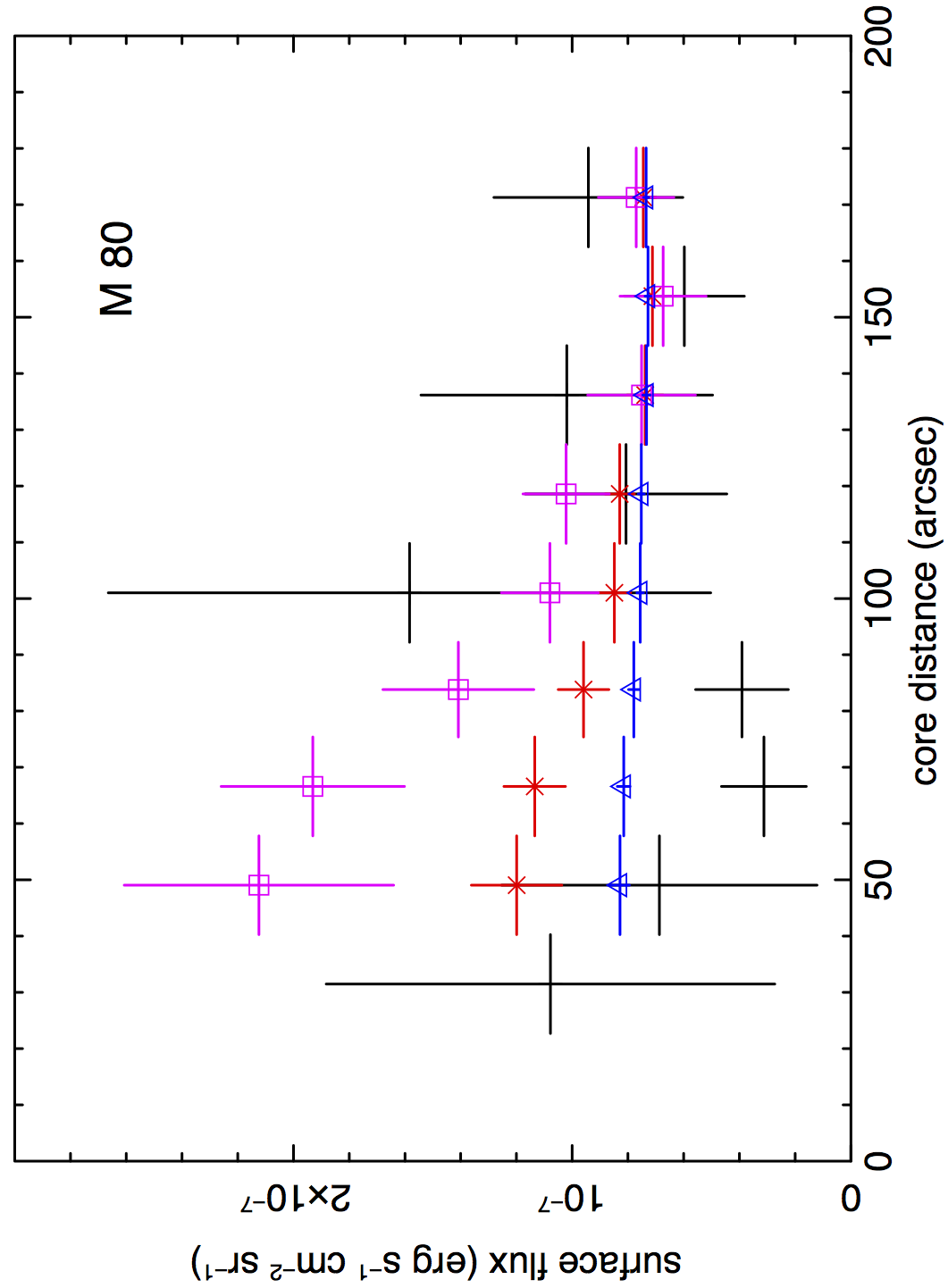}\end{turn}}
  \caption{Radial dependence of the surface flux (1--7~keV) above the NXB for four of the six GCs considered in this study (as labeled in the individual figures). 
  The different colors and markers show the measured surface flux (crosses, black) and the predictions based on an IC model (squares, magenta), a model assuming the same luminosity as \terzan\ (stars, red) and a model assuming a connection to msPSRs (triangles, blue). 
  The drawn uncertainties are the same as given in Table~\ref{tab-spectral-results}.
		}
  \label{fig-surface-brightness}
\end{figure*}

We fitted the NXB-subtracted spectrum from each annulus with an absorbed powerlaw model. 
Owing to lack of statistics in individual regions, the goal was only to reproduce the observed spectra reasonably well with a simple model to derive observed fluxes. 
For each GC we linked the hydrogen column density \nh\ for all eight rings, but left the photon index and the normalization free to vary independently. 
If more than one spectrum was available for a given annulus due to two or three contributing detector CCDs, we linked all spectral parameters, with a fixed ratio between the normalizations, corresponding to the relative size of the extraction areas on each CCD. 
As for \terzan , because of the limited statistics in the spectra from individual annuli, the absorbed powerlaw model reproduced the spectra sufficiently well with reduced $\chi^2$ values in the range of 1.1 to 1.4. 
We calculated the surface flux for each region by dividing the measured total flux by the effective extraction area, taking into account excluded regions, regions outside the FoV as well as bad columns and pixels. 
Table~\ref{tab-spectral-results} contains the spectral fitting results for those four GCs where significant excess emission above the NXB was detected. 
The radial evolution of the surface brightness is shown in Fig.~\ref{fig-surface-brightness}. 
Here we show the observed fluxes, because the absorbed powerlaw model is only a rough approximation of the spectral shape. 
A correction for absorption would require a fit of individual annuli with a more physically motivated model, which the limited statistics does not allow for. 
However, assuming that the foreground column density does not vary much within the regions, the observed and intrinsic fluxes can be expected to scale similarly.

The radial diffuse surface flux of M~80 seems to be peaked in the fourth bin at about 100$^{\prime\prime}$. 
We investigated this by visual examination of the raw images and smoothed versions in different energy bands for any point-like or extended but still compact source located in this annulus that might have escaped the detection algorithm somehow, but we did not find any. 
However, despite this difference, the fluxes in all annuli are still comparable within the uncertainties. 
Possible explanations for this increase are statistical fluctuations or a faint extended source, such as a background cluster of galaxies with a flux below the detection threshold. 

To investigate the nature of the observed diffuse X-ray emission in more detail, we fitted the combined spectrum extracted from the outer three regions with a more physically motivated model. 
In \terzan\ we found significant diffuse X-ray excess emission above the Galactic diffuse background only within the inner five annuli, whereas the combined spectrum from the outer three annuli was compatible with pure diffuse Galactic background. 
Even though we did not detect any significant dependence of the surface flux on the distance to the GC center, we still considered the inner five annuli as our source region of potential excess emission and the outer three annuli as the region containing predominantly background emission. 

Here we performed the same study as for \terzan\ in using a non-equilibrium ionization model (NEI in XSPEC) to describe the emission from diffuse hot Galactic gas. 
This model was already used by \citet[][henceforth E05]{2005ApJ...635..214E}, who performed deep \chandra\ observations of a typical ``empty" Galactic Plane region and in particular studied the diffuse emission after the removal of point sources in great detail. 
In contrast to the powerlaw fitting of individual annuli here we chose an energy interval of 0.5--10~keV to be able to directly compare our results to those obtained by E05. 
The column density and consequently also the expected amount of emitting gas is about an order of magnitude higher toward the region studied by E05 compared to the observations used here (\nh\ = 2.07\hcm{22} vs. \nh\ $\sim$ 1--2\hcm{21}, respectively). 
To account for the lack of statistics arising from shorter exposures and lower fluxes, we fixed most of the model parameters to the values given in Table 8 from E05 by leaving only the normalization and the temperature free to vary. 
To fit the spectra from the outer regions, we first started with only one temperature component and that already yielded good results for three GCs (M 62, NGC 6388, NGC 6541). 
In these cases the derived temperatures are comparable to the soft component in E05. 
For M 80 a second component was needed to describe the data and we used the hard component from E05 with a temperature fixed at their value ($kT = 5$\,keV). 
To compare the integrated observed surface fluxes to E05, we chose the same energy range (0.7--10.0\,keV). 
These results are also listed in in Table~\ref{tab-spectral-results}. 
As discussed in detail in sect.~\ref{sec-galactic-diffuse}, we conclude that the diffuse X-ray emission in all cases is compatible with the diffuse Galactic emission and no extra component above this background is found. 

\subsection{Upper limits}
\label{sec-upper-limits}
None of the six GCs shows signs of significant diffuse excess emission between 0.6 and 4.4 \rh\ above the NXB and Galactic diffuse background components. 
As mentioned in Sect.~\ref{sec-introduction}, this region could be interesting, because a significant diffuse signal was detected from the GC \terzan\ \citep{2010A&A...513A..66E}. 
To derive upper limits for diffuse excess emission from each GC, we used the total number of counts in the 1--7~keV band within the regions Ring1--5 and Ring0--5, respectively, again using the same exclusion regions as for the spectra. 
The first region was chosen such that the derived upper limits can be directly compared to models based on the level of diffuse X-ray excess flux from \terzan , which was measured from the respective region. 
The latter area (Ring0-5) also takes into account the additional region of Ring0, which was not possible for \terzan\ because of its large point source population compared to the GCs studied here. 
The upper limit from the Ring0-5 region would allow one to constrain models for diffuse X-ray flux that are not based on the \terzan\ measurement, but only rely on other already known parameters of the respective GC. 
Because in many cases the diffuse X-ray flux is expected to drop rapidly with increasing distance from the GC center, the additional area of Ring0, which is located closer to the GC cores, would produce tighter constraints on these models. 

Assuming that all the counts in the respective regions are due to background, we calculated the required number of excess counts to yield a detection with a statistical significance of 5\,$\sigma$. 
We converted the excess count-rate into a flux by assuming an absorbed powerlaw model with an \nh\ value from \citet{1990ARA&A..28..215D} for each GC position and a photon index of $-$0.9 \citep[as measured for \terzan , see][]{2010A&A...513A..66E}. 
To calculate flux upper limits for the total geometric area of Rings0--5 and Rings1--5 we scaled the values with the inverse of the coverage fraction, assuming a homogeneous flux distribution over the whole region. 
Note that the coverage fractions quoted in Table~\ref{tab-observations} also include excluded regions from point-like or diffuse sources as well as bad pixels and columns. 
Furthermore, we converted the flux upper limits into upper limits on the luminosity using the distances listed in the GC catalog of \citet{1996AJ....112.1487H}. 
Table~\ref{tab-upper-limits} contains the information on count statistics and the derived upper limits for each of the six GCs. 

\begin{table}[t]
\caption[]{GC upper limits}
\begin{center}
\begin{tabular}{llll}
\hline\hline\noalign{\smallskip}
\multicolumn{1}{l}{GC} &
\multicolumn{1}{l}{Counts$^{(1)}$} &
\multicolumn{1}{l}{Excess Rate$^{(2)}$} &
\multicolumn{1}{l}{UL $F_\mathrm{X}^{(3)}$} \\
\multicolumn{1}{l}{name} &
\multicolumn{1}{l}{} &
\multicolumn{1}{l}{(10$^{-3}$\,s$^{-1}$)} &
\multicolumn{1}{l}{(10$^{-13}$erg\,cm$^{-2}$\,s$^{-1}$)} \\
\noalign{\smallskip}\hline\noalign{\smallskip}
\multicolumn{4}{l}{\emph{Rings 1--5}} \\
M 62			& 13392	& 9.29 	& 1.1	   \\ 
NGC 6388		& 3749	& 6.85	& 0.89	    \\ 
NGC 6541		& 6681	& 9.10	& 1.1	    \\ 
M 28			& 49219	& 7.80	& 0.96	    \\ 
M 80			& 2591	& 5.24	& 0.63	    \\ 
NGC 6139		& 2197	& 13.0	& 1.7	   \\ 
\hline\noalign{\smallskip}
\multicolumn{4}{l}{\emph{Rings 0--5}} \\
M 62			& 14502	& 9.67	& 1.2	\\
NGC 6388		& 4121	& 7.18	& 0.93	\\
NGC 6541		& 7385	& 9.60	& 1.2	\\
M 28			& 55262	& 8.28	& 1.0	\\
M 80			& 2876	& 5.52	& 0.66	\\
NGC 6139		& 2386	& 13.5	& 1.7	\\
\noalign{\smallskip}\hline\noalign{\smallskip}
\end{tabular}
\label{tab-upper-limits}
\end{center}
$^{(1)}$Total number of counts in the inner five (top section) and inner six (bottom section) rings, respectively. 
$^{(2)}$Required excess rate for a 5\,$\sigma$ detection. 
$^{(3)}$Flux upper limit (1--7\,keV). Note that these upper limits on the flux are not yet corrected for the coverage fraction, but the luminosity upper limits quoted in Table~\ref{tab-model-predictions} are.  
\end{table}

\section{Discussion}
In this section the implications of the above analysis of diffuse X-ray emission from the six GCs is discussed. 
First, we compare the observed surface fluxes and spectra from the four GC where such an analysis was possible to what might be expected from pure Galactic diffuse emission from hot gas. 
The rest of the discussion will then evaluate three models for diffuse X-ray excess emission related to the GCs in the context of the derived upper limits. 
We also note here that the population of unresolved point sources below the detection threshold of the \chandra\ observations would contribute to any extended X-ray emission. 
Therefore, the quoted upper limits for diffuse emission may be considered as quite conservative in this sense. 
Because we particularly aim to determine whether the X-ray signal detected from \terzan\ is a general feature of GCs or whether it stands out as an exception, we use the flux measured from \terzan\ as a reference. 
Because there is no measurement of diffuse X-ray flux from \terzan\ corresponding to Ring0, we only used the upper limits derived from the region covered by Rings1--5 to compare with the model predictions in sections \ref{sec-mspsrs} to \ref{sec-same-lumi}. 

\subsection{Galactic diffuse emission}
\label{sec-galactic-diffuse}
Before the discussion of potential excess emission above the various background components, we first examine the NXB and diffuse Galactic background. 
Using the GTI filtering and the ``stowed" dataset, we are confident that the NXB has been removed as best as possible from the spectra. 
Instead of removing the Galactic diffuse background as well, we decided to model the emission because this component is much more dependent on the actual observation position and might vary in spectral shape as well as in intensity between different fields. 

In principle, one could compare the fluxes and temperatures derived from the NEI model fit with 
dedicated studies of Galactic diffuse emission, such as performed with \chandra\ by E05. 
As already mentioned in Sect.~\ref{sec-preparation-point-sources}, the column densities toward the regions  here and the field studied by E05 are about a factor of ten lower. 
Assuming that the amount of X-ray emitting gas is proportional to the amount of total interstellar material, we scaled the flux measured by E05 with the ratio of \nh . 
For all but one GCs (M\,80) the derived expected flux is about a factor of $\sim$5 below the measured value. 
This could be explained by the fact that most of the matter is also contributing to the absorption of the X-ray flux, which is strongly dependent on the spectral shape of the intrinsic spectrum and the line-of-sight geometry. 
Particularly for soft spectra ($kT < 1$\,keV) absorption has a very strong impact on the observed flux and it scales highly non-linear with \nh . 
We tested the relative influence of the intrinsic flux and the foreground absorption on the observed flux for our case based on the model parameters from E05. 
Increasing both \nh\ (from 10$^{21}$\,cm$^{-2}$ to 10$^{22}$\,cm$^{-2}$) and the intrinsic flux by a factor of ten the observed flux from the soft component ($kT = 0.3$\,keV) even decreases by 26\%, whereas the observed flux from the hard component ($kT = 5$\,keV) scales nearly as its intrinsic flux. 
This means that the flux expectation based on a scaling with \nh\ is also highly dependent on the relative contribution from both components. 

We found that for three of the four GCs (M\,62, NGC\,6388, NGC\,6541) the expected surface flux is compatible with the results from E05 taking the effects of absorption into account. 
Because a single (soft) component was sufficient to reproduce the spectra from M\,62 and NGC\,6388, we also only took this component into account. 
For M\,80, however, the observed surface flux is at the same level as found by E05 for a region with 17 times the \nh . 
This could be explained by the hard component contributing more than in the region studied by E05, which is also supported by the fact that only for this GC the addition of a 5~keV thermal component greatly increased the quality of the fit. 
However, the data for M\,80 have a much lower statistical quality compared to the region studied by E05 because of the shorter exposure time. 
Consequently, a more detailed study of the relative contribution and separate foreground absorption of the two components is not possible with the available data. 
Also, the diffuse thermal source in the vicinity of M\,80 detected by \citet{2007PASJ...59..727O} might be more extended than suggested by their image and could therefore contaminate our region of interest. 
In this case as well, only a dataset with a significantly longer exposure would be needed to better determine the extent of this relatively weak diffuse source. 

\begin{table}[t]
\caption[]{GC model predictions}
\begin{center}
\begin{tabular}{llllll}
\hline\hline\noalign{\smallskip}
\multicolumn{1}{l}{GC} &
\multicolumn{1}{l}{$d^{(1)}$} &
\multicolumn{1}{l}{UL$_{05}$ $L_\mathrm{X}^{(2)}$} &
\multicolumn{1}{l}{UL$_{15}$ $L_\mathrm{X}^{(3)}$} &
\multicolumn{1}{l}{IC$^{(4}$} &
\multicolumn{1}{l}{msPSRs$^{(5)}$} \\[0.1cm]
\multicolumn{1}{l}{name} &
\multicolumn{1}{l}{(kpc)} &
\multicolumn{4}{c}{(10$^{33}$erg\,s$^{-1}$)} \\
\noalign{\smallskip}\hline\noalign{\smallskip}
M 62			&	6.8		& 0.62	&	0.57	  &	15$\pm$2	&	 0.81$^{+0.65}_{-0.38}$	\\[0.2cm] 
NGC 6388		&	9.9		& 1.1	&	0.99	  &	19$\pm$3	&	 1.9$^{+2.7}_{-1.0}$	\\[0.2cm] 
NGC 6541		&	7.5		& 0.99	&	0.89	  &	8.3$\pm$1.2	&	 0.27$^{+0.21}_{-0.16}$	\\[0.2cm] 
M 28			&	5.5		& 0.57	&	0.51	  &	6.0$\pm$0.9	&	 0.46$^{+0.46}_{-0.25}$	\\[0.2cm] 
M 80			&	10.0	& 0.67	&	0.61	  &	6.4$\pm$0.9	&	 0.62$^{+0.57}_{-0.44}$	\\[0.2cm] 
NGC 6139		& 	10.1	& 2.3	&	2.1 	  &	7.2$\pm$1.0	&	 0.81$^{+0.46}_{-0.46}$	\\[0.1cm] 
\hline\noalign{\smallskip}
\end{tabular}
\label{tab-model-predictions}
\end{center}
$^{(1)}$Distance to the GC from \citet{1996AJ....112.1487H}. 
$^{(2)}$Luminosity upper limit (1--7\,keV) for rings 0--5. 
$^{(3)}$Luminosity upper limit (1--7\,keV) for rings 1--5. 
$^{(4)}$Luminosity prediction for an inverse Compton scenario. 
$^{(5)}$Luminosity prediction for a scenario based on millisecond pulsars. 
\end{table}

\subsection{Scenario 1: Synchrotron radiation from leptons produced by msPSRs}
\label{sec-mspsrs}
The TeV $\gamma$-ray signal detected from the direction of \terzan\ may be explained by IC emission from leptons accelerated either in pulsar wind nebulae (PWNe) surrounding individual msPSRs \citep{2007MNRAS.377..920B} or in colliding PWN shocks \citep{2010ApJ...723.1219C}. 
Following \citet{2008AIPC.1085..277V}, the IC emission in the TeV band is supposed to be accompanied by X-ray emission arising from synchrotron radiation from the same lepton population. 
However, in simple models the apparent offset of the TeV signal from the core of \terzan\ challenges this interpretation, because the IC signal would be expected to follow the morphology of the GC radiation field as well as that of the lepton sources, which would also be traced by the X-ray emission. 

Nevertheless, assuming leptons generated by msPSRs are indeed responsible for the X-ray and TeV emission seen from \terzan , this scenario can be tested for all six GCs. 
If the magnetic field strength is similar among all GCs, the X-ray luminosity then only scales with the number of msPSRs. 
The total number of msPSRs can be derived by assuming that the GeV signal detected by \fermi / LAT is caused by magnetospheric emission from individual pulsars \citep{2010A&A...524A..75A}. 
The total number of msPSRs then is proportional to the luminosity in the GeV range. 
Therefore, to estimate the X-ray luminosity for each of the six GCs we scaled the luminosity measured from \terzan\ with the ratio of the GeV luminosities. 
For this we used the results obtained by \citet{2010A&A...524A..75A} (M62, NGC 6388, M28) and \citet{2011ApJ...729...90T} (NGC 6541, M 80, NGC 6139), respectively. 
Our luminosity predictions can be found in Table~\ref{tab-model-predictions} (column labeled ``msPSRs"). 
The confidence intervals given in the table are based on the uncertainties of the \fermi\ and \chandra\ measurements. 

Only for the two GCs M 62 and NGC 6388 the model predictions lie clearly above the upper limits (UL$_{15}$). 
However, because of the relatively large error intervals of the \fermi\ measurements the values barely agree within the uncertainties. 
Therefore, we conclude that we cannot rule out a model based on msPSRs for any of our GCs, even though there are indications for discrepancies between upper limit and prediction for two of them. 
Because the dominant uncertainties come from the \fermi\ measurements, improved constraints on a msPSR scenario can be achieved from a GeV dataset with better statistics. 

\subsection{Scenario 2: IC emission from a GRB remnant}
\label{sec-ic}
The TeV $\gamma$-ray signal from \terzan\ might be associated to a recent short GRB event resulting from a neutron stars merger \citep[as discussed in][]{2011A&A...533L...5D}. 
In this scenario the TeV $\gamma$-rays would be produced in inelastic collisions of hadronic cosmic rays accelerated in the relativistic shock waves with ambient interstellar matter. 
Here the hard diffuse X-ray emission could arise from IC up-scattering by already cooled and mildly relativistic primary leptons accelerated at the GRB shock waves, if the lepton to hadron ratio is about 0.1.

Assuming that such a short GRB remnant is also present in our GCs, we calculated the expected X-ray flux based on the same lepton energy density required to explain the emission from \terzan . 
In this case, the X-ray luminosity would only scale with the energy density of the target photon field, where for GCs the main contribution comes from the starlight from their densely populated core regions. 
To calculate the expected X-ray luminosities for each of the six GCs, we converted the total measured diffuse X-ray flux of (5.5$\pm$0.8)\ergcm{-13} (1--7\,keV) from \terzan\ \citep{2010A&A...513A..66E} into a luminosity using the updated distance estimate of 5.9 kpc from \citep{2009Natur.462..483F} and scaled it with the ratio of the optical luminosity given by \citet{1996AJ....112.1487H}. 
The resulting X-ray luminosity for this GRB/IC scenario are given in Table~\ref{tab-model-predictions} (column labeled ``IC"). 
The uncertainties here are based on the 1\,$\sigma$ statistical uncertainty of the flux measurement from \terzan\ with \chandra . 
We also show the expected radial surface flux profile based on the measurements from \terzan . 
For this purpose we performed the same scaling as for the total source on the luminosities contained in each individual annulus. 
Then we calculated the expected surface flux from each annulus taking the distance to each GC and the surface area of the ring into account. 
For this model the predictions are shown as squares (magenta) data points in Fig.~\ref{fig-surface-brightness}. 
Errors only reflect the uncertainty of the measured radial profile from \terzan . 

It becomes clear from the values that an IC emission scenario from a putative GRB remnant with the above assumptions can be ruled out for every GC. 
The high values for the expected X-ray luminosities arise from the fact that all GC exhibit higher optical luminosities than Terzan 5. 
That leads to more efficient IC up-scattering in these systems. 

Two conclusions can be drawn here. 
Firstly, the non-detection of diffuse X-ray emission in our GC sample could result from the low rate
of GRBs in GCs.
As has been discussed in \citet[][and references therein]{2011A&A...533L...5D}, the rate of short
GRBs in the Milky Way is $\mathcal(O)$10$^{-4}(f_b^{-1}/100)^{-1}$ per year, where $1 \ll f_\mathrm{b} < 100$ is the beaming factor of short bursts.
Consequently, the probability to find two young GRB remnants in the Milky Way is low.
Secondly, if extended TeV emission in another GC were to be detected in the future and this emission can again be interpreted as a GRB remnant, this would point toward different remnant properties in the two systems.
Gamma-ray burst remnants could differ in the total energetics or the proton to electron ratio.

\subsection{Scenario 3: A source with the same luminosity as \terzan}
\label{sec-same-lumi}
The perhaps simplest assumption is that a source with the same X-ray luminosity as \terzan\ resides in each GC. 
Furthermore, in contrast to the previous scenario we assume that the X-ray signal does not arise from IC up-scattering of ambient photon fields and is also not dependent on any other property of the GC, or that the respective properties are similar to \terzan\ for the six GCs. 
In this case the X-ray luminosity upper limits only have to be compared to the luminosity of \terzan , which is (2.3$\pm$0.3)\ergs{33} \citep[see][assuming again a distance of 5.9 kpc]{2010A&A...513A..66E}. 
This scenario can be ruled out for every GC but NGC 6139, where the luminosity upper limit (UL$_{15}$) is below the expectation, but within the 1\,$\sigma$ statistical uncertainties. 
For this model we also calculated a predicted profile of the radial surface brightness (as described in Sect.~\ref{sec-same-lumi}),  which we show as stars (red) in Fig.~\ref{fig-surface-brightness}. 

\section{Conclusions}
\label{sec-conclusions}
Extended and diffuse X-ray emission with a very hard energy spectrum has recently been detected from \terzan .
The aim of this work was to search for extended and diffuse X-ray emission from other GeV-detected GCs to test whether \terzan\ stands out as an exception or if these properties can be found in a population of Galactic GCs.
We analyzed \chandra\ data of the six GCs M\,62, NGC\,6388, NGC\,6541, M\,28, M\,80 and NGC\,6139 and did not find any indications for diffuse excess emission on similar relative scales as seen from \terzan .

We used the resulting upper limits for the total diffuse X-ray luminosity to constrain three different scenarios, based on the assumption that the flux measured from \terzan\ arises from the respective emission scenario.
In these scenarios diffuse non-thermal X-ray emission is either arising from synchrotron radiation from the relativistic lepton population produced by the population of msPSRs, IC emission
from already cooled leptons accelerated along with the primary hadrons in a GRB remnant, or it was simply assumed that every GC hosts a source of diffuse X-ray emission with the same luminosity as the one detected in \terzan .

We were not able to rule out synchrotron emission in a msPSRs scenario for any of the six GCs. However, this process appears to be unlikely for NGC\,6388 and M\,62. 
On the other hand, a scenario based on a GRB remnant with the same properties as the putative GRB remnant in \terzan\ is ruled out for every GC. 
Lastly, the assumption that each GC hosts a diffuse X-ray source with the same luminosity as seen from \terzan\ is ruled out for all GCs but NGC\,6139.

Currently, \terzan\ is the only GC where a TeV $\gamma$-ray source is found in the vicinity of the cluster. 
The upper limits on diffuse X-ray emission from the other GCs will also be important for constraining the emission scenario for TeV $\gamma$-rays if additional GCs are detected in this energy band in the future.

\begin{acknowledgements}
The authors acknowledge the support from \'A-C Clapson, who contributed to our work on GCs in the past. 
This research has made use of data obtained from the Chandra Data Archive and software provided by the Chandra X-ray Center (CXC) in the application packages CIAO and ChIPS. 
\end{acknowledgements}

\bibliographystyle{aa}
\bibliography{citations_v2}

\end{document}